\documentclass[prd,aps,nofootinbib,floatfix,11 pt]{revtex4}
\usepackage{amssymb}
\usepackage{amsmath,graphicx,color,epsfig}

\begin{document}

\title{Dirac equation for quasi-particles in graphene and quantum field
theory of their Coulomb interaction}
\author{Riazuddin\footnote{riazuddin@ncp.edu.pk}}
\affiliation{National Centre for Physics,
Quaid-i-Azam University Campus,
Islamabad, 45320 Pakistan}

\begin{abstract}
There is evidence for existence of massless Dirac quasi-particles in
graphene, which satisfy Dirac equation in (1+2) dimensions near the so
called Dirac points which lie at the corners at the graphene's brilluoin
zone. We revisit the derivation of Dirac equation in (1+2) dimensions obeyed
by quasiparticles in graphene near the Dirac points. It is shown that parity
operator in (1+2) dimensions play an interesting role and can be used for
defining "conserved" currents resulting from the underlying Lagrangian for
Dirac quasi-particles in graphene which is shown to have $U_{A}(1)\times
U_{B}(1)$ symmetry. Further the quantum field theory (QFT) of Coulomb
interaction of 2D graphene is developed and applied to vacuum polarization
and electron self energy and the renormalization of the effective coupling $%
g $ of this interaction and Fermi velocity $v_{f}$ which has important
implications in the renormalization group analysis of $g$ and $v_{f}$.
\end{abstract}

\maketitle

Recent progress in the experimental realization of a single layer problem of
graphene has led to extensive exploration of electronic properties in this
system. Experimental and theoretical studies have shown that the nature of
quasiparticles in these two-dimensional system are very different from those
of the conventional two-dimensional electron gas (2DEG) system realized in
the semiconductor heterostructures. Graphene has a honeycomb lattice of
carbon atoms. The quasiparticles in graphene have a band structure in which
electron and hole bands touch at two points in the Brillouin zone. At these
Dirac points the quasiparticles obey the massless Dirac equation. In other
words, they behave as massless Dirac fermions leading to a linear dispersion
relation $\epsilon _{k}=vk$ (with the characteristic velocity $v\simeq 10^{6}$%
m/s). This difference in the nature of the quasiparticles in graphene from
conventional 2DEG has given rise to a host of new and unusual phenomena such
as anomalous quantum Hall effects and a $\pi $ Berry phase \cite{1, 2}.
These transport experiments have shown results in agreement with the
presence of Dirac fermions. The 2D Dirac-like spectrum was confirmed
recently by cyclotron resonance measurements and also by angle resolved
photoelectron spectroscopy (ARPEC) measurements in monolayer graphene \cite{3}%
. Recent theoretical work on graphene multilayer has also shown the
existence of Dirac electrons with a linear energy spectrum in monolayer
graphene \cite{4}.

In this work, we revisit the derivation of Dirac equation obeyed by
quasiparticles in graphene near the Dirac points. Since monolayer graphene
is a physical system (1+2) space-time dimensions this has important
consequences: Dirac matrices in odd number of space-time dimensions can have
two inequivalent representations and their role in obtaining the Dirac
equation for quasiparticles is emphasized here. Moreover, it is also shown
that the Lagrangian for the system has chiral $U_{A}\left( 1\right) \otimes
U_{B}\left( 1\right) $ symmetry. Furthermore, we have carried out a QFT
calculation of Coulomb interaction effects in graphene. The obtained results
are compared with those obtained in \cite{5}.The existence of
quasi-particles in graphene, which satisfy massless Dirac equation in (1+2)
dimensions near the so called Dirac points, which lie at the corners of
graphene's Brilluoin zone. This is indicated by energy dispersion around the
Dirac point given by \cite{6}
\begin{equation}
E_{\pm }\simeq \pm v_{f}\left\vert \mathbf{q}\right\vert +O(\left\vert
\mathbf{q}\right\vert ^{2})  \label{01}
\end{equation}%
which reminds one of relativistic energy-momentum relation for massless or
ultra relativistic particles although Fermi velocity is three hundred times
smaller than the speed of light. The derivation of such an equation given in
literature \cite{6, 7} is in the form
\begin{equation}
i\mathbf{\sigma }.\mathbf{\nabla }\psi =E\psi   \label{02}
\end{equation}%
where $\mathbf{\sigma }=(\sigma ^{1},$ $\sigma ^{2})$ and $\mathbf{\nabla }=(%
\frac{\partial }{\partial x^{1}},\frac{\partial }{\partial x^{2}}$ $),$
which has two dimensional form of the usual Dirac equation for two component
left-handed or right-hand massless fermions and not exactly in the form of
Dirac equation in (1+2) dimensions[see below]. The purpose of this paper is
to clarify such points and to discuss the subtleties of odd [in this case
three] space-time dimensions and develop a quantum field theory of Coulomb
interaction of 2D graphene.

In graphene for the hexagonal layer the unit cell contains two atoms A and
B, belonging to two sublattices. The lattice vectors are \cite{6}
\begin{equation}
\mathbf{a}_{1}=\frac{a}{2}(3,\sqrt{3}),\ \ \ \ \ \ \ \ \bigskip \mathbf{a}%
_{2}=\frac{a}{2}(3,-\sqrt{3})  \label{03}
\end{equation}%
where $a\simeq 1.42A^{0}$ is the carbon-carbon distance.

The Dirac points K and K$^{^{\prime }}$ have their position vectors in
momentum space:
\begin{equation}
\mathbf{K}=\frac{2\pi }{3a}(1,\sqrt{3}),\bigskip \ \ \ \ \ \ \ \mathbf{K}%
^{\prime }=\frac{2\pi }{3a}(1,-\sqrt{3})  \label{04}
\end{equation}%
In the tight binding approach, the Hamiltonian is given by \cite{6}
\begin{equation}
\mathcal{H}=-t\sum_{<i,j>}(a_{i}^{\dagger }b_{j}+h.c.)  \label{05}
\end{equation}%
where $a_{i}(a_{i}^{\dagger })$ are annihilation(creation) operators for an
electron in sub-lattice A with an equivalent definition for sub-lattice B. $%
t(\simeq 2.8$eV$)$is the nearest hopping energy (hopping between different
sub-lattices).

Introduce now Fourier transform of field operator:%
\begin{eqnarray}
a_{i} &=&\sum\limits_{k}a_{ki}e^{i\vec{k}.\vec{u}_{i}}  \nonumber \\
b_{j} &=&\sum\limits_{k}b_{kj}e^{i\vec{k}.\vec{u}_{j}}  \label{5}
\end{eqnarray}%
where $.\vec{u}_{i}($or $\vec{u}_{j})$ may be taken any of two independent
lattice points \cite{2}. $\mathbf{a}_{1}$ and $\mathbf{a}_{2}$ which may
taken as in Eq. (\ref{03}). Then Eq. (\ref{05}) gives%
\[
\mathcal{H}=\sum\limits_{k}\left(
\begin{tabular}{ll}
$a_{ki}^{\dagger }$ & $b_{kj}^{\dagger }$%
\end{tabular}%
\right) \mathcal{H}_{D}\left(
\begin{tabular}{l}
$a_{ki}$ \\
$b_{kj}$%
\end{tabular}%
\right)
\]

where
\[
\mathcal{H}_{D}=\left(
\begin{array}{ll}
0 & \mathcal{H}_{AB} \\
\mathcal{H}_{AB}^{\ast } & 0%
\end{array}%
\right)
\]%
with
\begin{equation}
\mathcal{H}_{AB}=-t[e^{i\mathbf{k}.\mathbf{a}_{1}}+ie^{i\mathbf{k}.\mathbf{a}%
_{2}}]  \label{06}
\end{equation}%
Let us expand around the Dirac point $K^{^{\prime }}.$
\[
\mathbf{k}=\mathbf{K}^{^{\prime }}+\mathbf{q}
\]%
keep terms linear in $\left\vert \mathbf{q}\right\vert ,$%
\begin{equation}
\mathcal{H}_{AB}=\frac{3at}{2}[iq_{x}+q_{y}]  \label{07}
\end{equation}%
This gives
\begin{eqnarray}
\mathcal{H}_{D} &=&\left(
\begin{array}{ll}
0 & iq_{x}+q_{y} \\
-iq_{x}+q_{y} & 0%
\end{array}%
\right)   \nonumber \\
&=&v_{f}[-q^{1}\sigma ^{2}+q^{2}\sigma ^{1}]  \label{08}
\end{eqnarray}%
where $v_{f}=\frac{3at}{2}$ is the Fermi velocity, $\sigma ^{1}$ and $\sigma
^{2}$ are Pauli matrices, $q^{i}=(q^{1},q^{2})$, in relativistic notation: $%
q^{\mu }=(q^{0},q^{1},q^{2}).$ This does not give $\mathbf{\sigma }.\mathbf{q%
}$ as implied in Eq. (\ref{02}). The Eq. (\ref{08}) can be put in the form
\begin{eqnarray}
\mathcal{H}_{D} &=&v_{f}[i\sigma ^{3}\sigma ^{1}q^{1}+i\sigma ^{3}\sigma
^{2}q^{2}]  \nonumber \\
&=&\gamma ^{0}(\mathbf{\gamma }.\mathbf{q}v_{f})  \label{09}
\end{eqnarray}%
where $\gamma ^{0}=\sigma ^{3},$ $\gamma ^{1}=i\sigma ^{1},$ $\gamma
^{2}=i\sigma ^{2}$ are Dirac matrices in (1+2) dimensions. The above is the
Dirac Hamiltonian for massless fermion in (1+2) dimensions near the Dirac
point $K^{^{\prime }}.$ The corresponding Dirac equation is
\begin{equation}
\mathcal{H}\psi =E\psi _{\pm },\bigskip E=\pm v_{f}\left\vert \mathbf{q}%
\right\vert   \label{10}
\end{equation}%
or
\begin{equation}
i\frac{\partial \psi }{\partial t}=\gamma ^{0}(\mathbf{\gamma }(-i\mathbf{%
\nabla }v_{f})\psi   \label{11}
\end{equation}%
Writing $\partial _{0}=\frac{1}{v_{f}}\frac{\partial }{\partial t}$, we can
write the Eq (\ref{10}) in covariant form
\[
i(\gamma ^{0}\partial _{0}+\gamma ^{1}\partial _{1}+\gamma ^{2}\partial
_{2})\psi =0
\]%
or
\begin{equation}
i(\gamma ^{\mu }\partial _{\mu })\psi =0  \label{12}
\end{equation}%
which is the Dirac equation in (1+2) dimensions for a massless fermion \cite%
{8}.

It is important to remark that if we expand around the Dirac point K, we
obtain
\begin{eqnarray}
\mathcal{H}_{D} &=&v_{f}[-q^{1}\sigma ^{2}-q^{2}\sigma ^{1}]  \nonumber \\
&=&v_{f}[i\sigma ^{3}\sigma ^{1}q^{1}\sigma ^{1}-i\sigma ^{3}\sigma
^{2}q^{2}]  \label{13}
\end{eqnarray}%
Now it is known \cite{8} that in 3 space-time dimensions there exists two
inequivalent representations for $\gamma $-matrices [this is true for any
odd number of space-time dimensions]:
\begin{eqnarray}
\gamma ^{0} &=&\sigma ^{3},\bigskip \gamma ^{1}=i\sigma ^{1},\gamma
^{2}=i\sigma ^{2}  \nonumber \\
\gamma ^{0} &=&\sigma ^{3},\bigskip \gamma ^{1}=i\sigma ^{1},\gamma
^{2}=-i\sigma ^{2}  \label{14}
\end{eqnarray}

We have used the first of these representations for the expansion around the
Dirac point K$^{^{\prime }}$. We take the second representation for the
expansion around the Dirac point K, whcih is obtained from K$^{^{\prime }}$
by the parity operation, defined by the matrix \cite{8}
\begin{equation}
\Lambda =\left(
\begin{array}{lll}
1 & 0 & 0 \\
0 & 1 & 0 \\
0 & 0 & -1%
\end{array}%
\bigskip \right)   \label{15}
\end{equation}%
so that det$\Lambda =-1.$ Thus we see that under the parity operation
\[
q^{1}\longleftrightarrow q^{1}\bigskip ,q^{2}\longleftrightarrow
-q^{2}\bigskip \text{ and }\mathcal{H}_{K}\longleftrightarrow \mathcal{H}%
_{K^{\prime }}
\]

Taking the two representations mentioned above into account we can write the
parity conserving Lagrangian \cite{8} as (it is instructive to put mass term
which can be put equal to zero when necessary)
\[
\mathcal{L}=\overline{\psi }_{+}(i\partial -mv_{f}^{2})\psi _{+}+\overline{%
\psi }_{-}(i\widetilde{\partial }-mv_{f}^{2})\psi _{-}
\]%
where
\begin{eqnarray}
\partial  &=&\gamma ^{0}\partial _{0}+\gamma ^{1}\partial _{1}+\gamma
^{2}\partial _{2}  \nonumber \\
\widetilde{\partial } &=&\gamma ^{0}\partial _{0}+\gamma ^{1}\partial
_{1}-\gamma ^{2}\partial _{2}  \label{16}
\end{eqnarray}%
Parity operation takes the solutions in one representation to the other:
\begin{eqnarray}
\psi _{+}^{p}(x^{p}) &=&-\eta _{p}\psi _{-}(x)  \nonumber \\
\psi _{-}^{p}(x^{p}) &=&-\eta _{p}\psi _{+}(x)  \label{17}
\end{eqnarray}%
where $x^{p}=(x^{0},x^{1},-x^{2})$. It is convenient to transform to new
fields \cite{8}
\begin{eqnarray}
\psi _{A} &=&\psi _{+}  \nonumber \\
\psi _{B} &=&i\gamma ^{2}\psi _{-}  \label{18}
\end{eqnarray}%
The Lagrangian (\ref{16}) can then be written as \cite{8}
\begin{equation}
\mathcal{L}=\overline{\psi }_{A}(i\gamma ^{\mu }\partial _{\mu
}-mv_{f}^{2})\psi _{A}+\overline{\psi }_{B}(i\gamma ^{\mu }\partial _{\mu
}+mv_{f}^{2})\psi _{B}  \label{19}
\end{equation}%
where under partiy operation now
\begin{eqnarray}
\psi _{A}^{p}(x^{p}) &=&-i\eta _{p}\gamma ^{2}\psi _{B}(x)  \nonumber \\
&=&\eta _{p}\sigma ^{2}\psi _{B}(x)  \nonumber \\
\psi _{B}^{p}(x^{p}) &=&-i\eta _{p}\gamma ^{2}\psi _{A}(x)  \nonumber \\
&=&\eta _{p}\sigma ^{2}\psi _{A}(x)  \label{20}
\end{eqnarray}

It may be noted that the Lagrangian (\ref{19}) is invariant under two
independent transformations
\begin{equation}
\psi _{A}\rightarrow e^{i\alpha _{A}}\psi _{A},\bigskip \psi _{B}\rightarrow
e^{i\alpha _{B}}\psi _{B}  \label{23}
\end{equation}%
where $\alpha _{A}$ and $\alpha _{B}$ are real, and has thus $%
U_{A}(1)\otimes U_{B}(1)$ symmetry. The corresponding conserved currents are
\begin{eqnarray}
J_{A}^{\mu } &=&\overline{\psi }_{A}\gamma ^{\mu }\psi _{A}  \nonumber \\
J_{B}^{\mu } &=&\overline{\psi }_{B}\gamma ^{\mu }\psi _{B}  \label{24}
\end{eqnarray}

When can form even (odd) combinations corresponding to "vector" ("axial
vector") under parity
\begin{equation}
J_{\pm }=(\overline{\psi }_{A}\gamma ^{\mu }\psi _{A}\pm \overline{\psi }%
_{B}\gamma ^{\mu }\psi _{B})  \label{023}
\end{equation}

We can develop quantum field theory for Coulomb interaction of 2D graphene
in analogy with QED (for another approach see \cite{7, 9}). The ``free''
Hamiltonian as implied by Eq. (\ref{11}) is
\begin{eqnarray}
\mathcal{H}_{o} &=&v_{f}\int d^{2}r\psi ^{\dagger }(\mathbf{r})\gamma ^{o}%
\mathbf{\gamma .}(-i\mathbf{\nabla })\psi (\mathbf{r})  \nonumber \\
&=&v_{f}\int d^{2}r\overline{\psi }(\mathbf{r})\mathbf{\gamma .}(-i\mathbf{%
\nabla })\psi (\mathbf{r})  \label{25}
\end{eqnarray}
The instantaneous Coulomb interaction in 2D graphene is
\begin{equation}
\mathcal{H}_{I}=\frac{e^{2}}{2}\int d^{2}x_{1}d^{2}x_{2}\frac{n(t,\mathbf{x}%
_{1})n(t,\mathbf{x}_{2})}{4\pi \left| \mathbf{x}_{1}-\mathbf{x}_{2}\right| }
\label{26}
\end{equation}
where $n(t,\mathbf{x}_{1})=\psi ^{\dagger }(t,\mathbf{x}_{1})\psi (t,\mathbf{%
x}_{1})=\overline{\psi }(t,\mathbf{x}_{1})\gamma ^{o}\psi (t,\mathbf{x}%
_{1}). $ This gives the scattering matrix element between four fermions
\begin{equation}
S_{fi}^{coulomb}=-ie^{2}\frac{1}{2}\int dt\int d^{2}x_{1}\int d^{2}x_{2}%
\frac{<f|n(t,\mathbf{x}_{1})n(t,\mathbf{x}_{2})|i>}{4\pi \left| \mathbf{x}%
_{1}-\mathbf{x}_{2}\right| }  \label{27}
\end{equation}
The integral can be written as $\int d^{D}x_{1}d^{D}x_{2}\delta
(t_{2}-t_{1}) $ where $D=3$ in (1+2) dimensions. Expanding $\psi ^{\prime }$%
s into creation and annihilation operators, one finally obtains \cite{10}
\begin{equation}
iT_{fi}=\frac{1}{4\pi v_{f}}[\overline{u}(p_{2}^{\prime })(-ie\gamma ^{\mu
})u(p_{2})\frac{i\eta _{\mu }\eta _{\nu }}{2[-q^{2}+(q.\eta )^{2}]^{\frac{1}{%
2}}}\overline{u}(p_{1}^{\prime })(-ie\gamma ^{\mu })u(p_{1})-\text{crossed
term}]  \label{28}
\end{equation}
where $\frac{1}{2\left| \mathbf{q}\right| }$, $\left| \mathbf{q}\right| =%
\left[ -q^{2}+(q.\eta )^{2}\right] ^{1/2}$is Fourier transform of $\frac{1}{%
4\pi \left| \mathbf{x}_{1}-\mathbf{x}_{2}\right| }$ in two space dimensions,
$\eta ^{\mu }=(1,0,0),$ so that $\eta _{\mu }\gamma ^{\mu }=\gamma ^{o}$ and
longitudinal photon momentum may be taken as $l^{\mu }\equiv 2(q^{\mu }-\eta
.q\eta ^{\mu })$, $\mu =0,1,2$; so that $(-l^{2})^{1/2}=2\left| \mathbf{q}%
\right| .$

Thus we may write Feynman rules for 2D coulomb interaction in analogy with
QED as follows:

(i) Vertex factor$:$ $-ie\gamma ^{\mu },$

(ii) Internal lines

Photon (``longitudinal'') line$:\frac{i\eta _{\mu }\eta _{\nu }}{%
2[-q^{2}+(q.\eta )^{2}]^{\frac{1}{2}}}$

(iii) Spin$\frac{1}{2}$ massless fermion$:\frac{i}{\not{p}}$

$\not{p}=\gamma ^{\mu }p_{\mu }$ $\mu =0,1,2,$ this follows from the
Lagrangian given in Eq. (\ref{19}).

As an application of these rules, we calculate vacuum polarization II$\left(
q^{2}\right) =-\eta _{\mu }\eta _{\nu }$ II$^{\mu \nu }(q),$ which arises
from photon self energy due to fermion loop. It is given by
\begin{equation}
-i\text{II}^{\mu \nu }(q)=2(-1)\frac{1}{v_{f}}\int \frac{d^{D}l}{(2\pi )^{D}}%
Tr[(-ie\gamma ^{\mu })\frac{i}{\not{l}+i\epsilon }(-ie\gamma ^{\nu })\frac{i%
}{\not{l}+2(\not{q}-q.\eta )\not{\eta}+i\epsilon }],  \label{29}
\end{equation}
where the factor of $2$ arises due to the two terms in Lagrangian (\ref{22}%
). The integration will be done by using Feynman parametrization and
dimensional regularization. Taking the trace in Eq. (29), using the Feynman
parametrization and making the shift $l\rightarrow l-2qx+2(q-\eta )\eta x$,
we obtain
\begin{equation}
-i\text{II}^{\mu \nu }(q)=-2i\frac{D+1}{2}\frac{e^{2}}{v_{f}}%
\int_{0}^{1}dx\int \frac{d^{D}x}{\left( 2\pi \right) D}\left[ \left( 2l^{\mu
}l^{\nu }-g^{\mu \nu }l^{2}\right) +g^{\mu \nu }L+x\left( x-1\right) A^{\mu
\nu }\right] \frac{1}{\left( l^{2}-L\right) ^{2}}  \label{30}
\end{equation}
where $(\frac{D+1}{2})$ arises from the trace for odd space time dimensions,
$L=4[(q.\eta )^{2}-q^{2}]$ and
\begin{equation}
A^{\mu \nu }=-8g^{\mu \nu }((q.\eta )^{2}-q^{2})-8q^{\mu }q^{\nu }+8q.\eta
(q^{\mu }\eta ^{\nu }+\eta ^{\nu }q^{\mu }-(q.\eta )\eta ^{\mu }\eta ^{\nu
}).  \label{31}
\end{equation}
The contribution from singular terms $\left( 2l^{\mu }l^{\nu }-g^{\mu \nu
}l^{2}\right) $ combines to give $-g^{\mu \nu }(1-D/2)\Gamma
(1-D/2)L=-g^{\mu \nu }\Gamma (2-D/2)L$, i.e. finite answer, which however
cancels with the contribution from the second term in parenthesis in Eq. (%
\ref{30}). The net result is
\begin{equation}
\text{II}^{\mu \nu }(q^{2})=\frac{2e^{2}}{v_{f}}\frac{1}{(2\pi )^{\frac{D}{2}%
}}(\frac{D+1}{2})\Gamma (2-\frac{D}{2})\int_{0}^{1}dx\sqrt{x(1-x)}\frac{%
A_{\mu \nu }}{2\Delta }  \label{32}
\end{equation}
where $\Delta =[(q.\eta )^{2}-q^{2}]^{\frac{1}{2}}=\left| \mathbf{q}\right| $.

Using $D=3$, we finally get
\begin{eqnarray}
\text{II(}q^{2}\text{)} &=&-\eta _{\mu }\eta _{\nu }\text{II}^{\mu \nu }=%
\frac{e^{2}}{4v_{f}}(\Delta )  \nonumber \\
&=&\frac{e^{2}}{4\pi v_{f}}\left| \mathbf{q}\right| \pi  \nonumber \\
&=&g\pi \left| \mathbf{q}\right| ,  \label{33}
\end{eqnarray}
in agreement with the known result [6, 7], where $g=\frac{e^{2}}{4\pi v_{f}}$
is the dimensionless [in units $\hbar=1$] effective coupling constant.

We now discuss how the vacuum polarization renormalizes the interaction
coupling constant. For this purpose we consider the response of charged
fermion to an externally applied field [Coulomb potential in two space
dimensions in momentum space is$A_{o}=-\frac{e}{2\left| \mathbf{q}\right| }%
=\eta _{\mu }A^{\nu },$ where $A^{\nu }=(-\frac{e}{2\left| \mathbf{q}\right|
},$ $\mathbf{0})]$, namely
\begin{equation}
\overline{u}(-ie\gamma ^{\nu })uiA_{\nu }=\overline{u}\left( -i\gamma
^{0}\right) u\left( -e^{2}/2\left| \mathbf{q}\right| \right)  \label{34}
\end{equation}
which is modified to
\begin{eqnarray}
&&\overline{u}\left( -ie\gamma ^{0}\right) uA_{0}\left[ \frac{1}{1-\eta
_{\lambda }\eta \rho \text{II}^{\lambda \rho }/\left( 2\Delta \right) }%
\right] A_{\nu }  \nonumber \\
&=&\overline{u}\left( -i\gamma ^{0}\right) u\left( -e^{2}/2\left| \mathbf{q}%
\right| \right) 1/\left[ 1+\text{II(}q^{2}\text{)}/2\Delta \right]  \nonumber
\\
&=&\overline{u}\left( -i\gamma ^{0}\right) u\left[ -g\frac{4\pi v_{f}}{%
2\left| \mathbf{q}\right| }\frac{1}{1+\frac{g\pi }{2}}\right]  \label{35}
\end{eqnarray}
where we have used Eq. (\ref{33}). Thus the renormalized $g$, often written
as g$_{sc}$ \cite{11} is
\begin{equation}
g_{sc}=\frac{g}{1+\frac{g\pi }{2}}  \label{36}
\end{equation}
The implications of this result are discussed in [6, 7]. It is instructive
to calculate fermion self energy which is given by
\begin{equation}
-i\Sigma (p)=\frac{1}{v_{f}}\int \frac{d^{3}l}{(2\pi )^{3}}(-ie\gamma ^{\mu
})\frac{i}{\not{l}}(-ie\gamma ^{\nu })\frac{i\eta _{\mu }\eta _{\nu }}{%
2\left| \mathbf{l}-\mathbf{p}\right| }  \label{37}
\end{equation}
After making the Wick rotation $l^{0}=il_{E}^{0}$ ($E$ for Euclidean matric)
and carrying out the $l_{E}^{0}$ integration (which occurs only in fermion
propagator), which gives $\pi \frac{1}{\left| \mathbf{l}\right| }$, the rest
of integration is in $D=2$ dimension. Using Feynman parametrization
\begin{equation}
\frac{1}{a^{1/2}b^{1/2}}=\frac{1}{\pi }\int_{0}^{1}x^{-1/2}(1-x)^{-1/2}\frac{%
1}{ax+b(1-x)},  \label{38}
\end{equation}
and the dimensional regularization one obtains
\begin{eqnarray}
-i\Sigma (p) &=&-\frac{ie^{2}}{16\pi v_{f}}\int_{0}^{1}x^{-1/2}\left(
1-x\right) ^{-1/2}\int \frac{d^{D}l}{\left( 2\pi \right) ^{D}}\frac{-\not{%
\eta}\not{p}\not{\eta}x}{\left( \mathbf{l}^{2}-L\right) }  \nonumber \\
&=&-\frac{ie^{2}}{4\pi v_{f}}\int_{0}^{1}x^{1/2}\left( 1-x\right) ^{-1/2}%
\frac{1}{\left( 4\pi \right) ^{D}}\frac{\Gamma \left( 1-D/2\right) }{\Gamma
\left( 1\right) }\left( \frac{1}{L}\right) ^{1-D/2}  \label{39} \\
&=&-\frac{ie^{2}}{32\pi v_{f}}(-\not{\eta}\mathbf{\gamma }\cdot \mathbf{p}%
\not{\eta})\ln \frac{\Lambda ^{2}}{\left| \mathbf{p}\right| ^{2}}  \nonumber
\end{eqnarray}
where we have used
\[
\frac{1}{(4\pi )^{(D/2-1)}}\frac{\Gamma (1-D/2)}{\Gamma (1)}\left( \frac{1}{L%
}\right) ^{1-D/2}=\left[ \frac{2}{\varepsilon }-(\ln L+\gamma -\ln 4\pi )%
\right]
\]
with $L=\left| \mathbf{p}\right| ^{2}x(1-x)$, $\frac{2}{\varepsilon }$
signifies the ultraviolet log divergence: $\ln \Lambda ^{2}$.

Noting that
\begin{eqnarray}
-\not{\eta}\mathbf{\gamma }\cdot p\not{\eta} &=&\mathbf{\gamma }\cdot
\mathbf{p}=-\not{p}+\gamma ^{0}p^{0}  \nonumber \\
\Sigma (p) &=&\frac{g}{4}\left[ -\not{p}+\gamma ^{0}p^{0}\right] \ln \frac{%
\Lambda }{\left\vert \mathbf{p}\right\vert }  \label{40}
\end{eqnarray}%
where $p^{0}=v_{f}\left\vert \mathbf{p}\right\vert $ is the energy. The
usual interpretation of this result is that coefficient of second term in
parenthesis gives the radiative correction to energy
\begin{equation}
E=v_{f}\left\vert \mathbf{p}\right\vert Z_{2}^{-1}  \label{R11}
\end{equation}%
where
\begin{equation}
Z_{2}^{-1}=1+\frac{g}{4}\ln \frac{\Lambda }{\left\vert \mathbf{p}\right\vert
}  \label{42}
\end{equation}%
The coefficient of $\not{p}$ gives the renormalization of the electric
charge, $e\rightarrow Z_{2}e,$which however is canceled by the
corresponding contribution from the vertex part by the use of\ Ward
Identity. The vacuum polarization correction is finite and renormalize $g$
to $g_{sc}$ as given in Eq. (\ref{36}). The result (\ref{R11}) is
interpreted as renormalization of the Fermi velocity $v_{f}$ \cite{7, 9, 11}$%
.$ Thus [putting $\left\vert \mathbf{p}\right\vert =q$ ]
\begin{equation}
v_{f}(q)=v_{f0}Z_{2}^{-1}=v_{f0}+v_{f0}\frac{g}{4}\ln \frac{\Lambda }{q}.
\label{43}
\end{equation}%
We remove the cut-off and unrenormalized $v_{f0}$ by subtraction at $q=q_{0}$%
:
\begin{equation}
v_{f}(q)-v_{f}(q_{0})=\frac{g}{4}v_{f0}\ln \frac{q_{0}}{q}=\frac{g}{4}%
v_{f}(q_{0})\ln \frac{q_{0}}{q}+O\left( g^{2}\right)  \label{44}
\end{equation}%
Since
\[
g=\frac{e^{2}}{4\pi v_{f}}
\]%
this gives \cite{11}
\begin{equation}
g(q)=g_{0}\left[ 1+\frac{g_{0}}{4}\ln \frac{q_{0}}{q}\right] ^{-1}=\left[
g_{0}^{-1}+\frac{1}{4}\ln \frac{q_{0}}{q}\right] ^{-1}  \label{45}
\end{equation}%
where $g_{0}=g(q_{0})$. One may take $q_{0}$ as the inverse of the lattice
constant [for our case $q_{0}=4nm^{-1}$]. An application of this result has
been discussed in \cite{5}.

In summary we have clarified the derivation of Dirac equation for
quasi-particles in graphene in (1+2) dimensions near the Dirac points $%
K^{^{\prime }}$ and $K$. The role of two inequivalent representations of
Dirac matrices in (1+2) dimensions [a property of odd number of space-time
dimensions] visa vis parity operation is emphasized. It is shown that the
underlying Lagrangian for quasi-particles in graphene has chiral $%
U_{L}(1)\otimes U_{R}(1)$ symmetry. Further Feynman rules for QFT of Coulomb
interaction of 2D graphite have been given and applied to the vacuum
polarization and renormalization of effective Coulomb interaction constant
and electron self energy which has important implication in the
renormalization group analysis of $g$ and $v_{f}$ \cite{11}.

The author would like to thank Ansar Fayyazuddin for introducing him to this
subject by giving him the reference for the review article \cite{6} . He
would also like to express his deep appreciation to Kashif Sabeeh for
reading this manuscript and making some useful suggestions. Author would
also like to acknowledge the hospitality of King Fahad University of
Patroleum and Minerals, Dhahran where a part of the work was completed.

\end{document}